\documentstyle[prb,preprint,aps]{revtex}
\begin{document}

\draft

\title{Correlations in Two-Dimensional Vortex Liquids}

\author{Jun Hu and A. H. MacDonald}

\address{Department of Physics, Indiana University, Bloomington, IN 47405}

\date{\today}
\maketitle

\begin{abstract}

We report on a high temperature perturbation expansion
study of the superfluid-density spatial correlation function
of a Ginzburg-Landau-model superconducting film in a magnetic field.
We have derived a closed form which expresses the contribution to
the correlation function from each graph of the perturbation
theory in terms of the number of Euler paths around
appropriate subgraphs.  We have enumerated all graphs appearing
out to 10-th order in the expansion and
have evaluated their contributions to the correlation function.
Low temperature correlation functions, obtained using Pad\'{e}
approximants, are in good agreement with Monte Carlo simulation
results and show that the vortex-liquid becomes strongly correlated
at temperatures well above the vortex solidification temperature.

\end{abstract}


\narrowtext

\section{Introduction}

Because of the combination of high transition temperatures,
strong anisotropy, and short coherence lengths which occurs in high
temperature superconductors, strong thermal fluctuations are present
over a wide temperature interval in these materials.
Thermal fluctuations are especially important in a magnetic field
where they are responsible for the melting
of the Abrikosov\cite{abrikosov} vortex lattice at temperatures below
the mean-field critical temperature giving rise to
a vortex liquid state\cite{liquid}.
In this paper we report on a study of correlations in the vortex liquid
in the extreme anisotropy limit of decoupled layers.
We have evaluated the leading terms in
the high temperature perturbation expansion of the superfluid density
spatial correlation function for the Ginzburg-Landau model of a
superconductor film.

The thermodynamics of this system is unusual
because of Landau quantization of the order parameter
fluctuations\cite{leeshenoy}; our calculations are carried out
within the lowest Landau level approximation in which
only the lowest gradient energy fluctuations are retained.
This approximation is valid near the mean-field transition
temperature and is ordinarily valid throughout the vortex
liquid state,
although fluctuation effects may be strong enough in some
high temperature superconductors to drive the solidification
transition outside of its range of validity.  High-temperature
perturbation expansions for the free energy of this model,
even when evaluated to high order\cite{ref:1,ref:2}, exhibit little
evidence of the Abrikosov vortex lattice state which is expected
to occur at low temperatures.  Recent Monte-Carlo
simulations\cite{ref:3,kato,ref:4,moore}, on the other hand,
generally obtain results indicative of a
weak first order phase transition between the two states.
(See however Ref.~[\onlinecite{moore}] where a different
conclusion is reached.)
We find that the perturbation expansion for the superfluid
density correlation function,
unlike that for the average superfluid density studied
by previous workers~\cite{ref:1,ref:2}, provides clear evidence
of strong correlations in the vortex liquid which presage
the appearance of an ordered state at low temperatures.

The paper is organized as following.
In Section II we introduce the quantity we study,
 $\langle |\Delta (\vec q)|^2 \rangle$ which
is proportional to the Fourier transform of the superfluid-density
spatial correlation function.
In Section III we outline the high temperature perturbation
expansion for
this quantity and present our closed form result for the
contribution to it from individual diagrams which appear in the
expansion.
In Section IV we discuss our evaluation of all terms out to
tenth order in this expansion and discuss some comparisons with
expansions for the free energy performed by earlier workers
which allow us to check our results.  Extrapolations of our
finite order results to low temperatures
using  Pad\'{e} approximants are presented and
compared with Monte Carlo simulations.
Section V contains a brief summary.

\section{Superfluid Density Correlation Function}

The free energy per unit area of the Ginzburg-Landau
model superconducting film in a perpendicular magnetic field is
given by
\begin{equation}
f[\Psi]  = \alpha (T) \vert \Psi \vert^2 + \frac{\beta}{2}
\vert \Psi \vert^4
+ \frac{1}{2m}\vert (-i \hbar  \nabla - 2e \vec A)\Psi \vert^2
\label{eq:1}
\end{equation}
where $\Psi ( \vec r)$ is the order parameter.  $\Psi (\vec r)$
is proportional to the wavefunction for the center of mass of the
Cooper pairs\cite{ajp}
and $\vec B = \nabla \times \vec A $.   (We employ the Landau
gauge so that $\vec A = (0, Bx, 0)$.)
The quadratic terms in Eq.~(\ref{eq:1}) are minimized by
order-parameters which correspond to a lowest Landau level
(LLL) wavefunction for the Cooper pairs.  In using Eq.~(\ref{eq:1})
we are neglecting screening so that $\vec B$ is the
external magnetic field.  This approximation is valid for type-II
superconductors near the upper critical field, the regime of
interest in this article.
The mean-field-theory superconducting instability occurs
when the quadratic terms in Eq.~(\ref{eq:1})
become negative for the LLL, {\it i.e.}
at $T_c^{MF}$ where $\alpha_H(T_c^{MF})=0$. ($\alpha_H \equiv \alpha
+ \hbar e B / m^*$. ) In the
LLL approximation we assume that fluctuations
in higher Landau level channels can be neglected or at
least absorbed in a
renormalization of $\alpha (T)$\cite{ref:1,ref:3,ikeda,vecpot}.
Then the order parameter $ \Psi (\vec r) $ can be expanded as
\begin{equation}
	\Psi (\vec r) = \sum_p C_p (L_y)^{-1/2}(\frac{2eB}{\pi
\hbar})^{1/4}
e^{ipy} e^{-\frac{eB}{\hbar}(x-\frac{\hbar p}{2eB})^2}
\label{eq:2}
\end{equation}
where the number of terms in the sum over $p$ is
$ N_\phi = L_xL_y/2\pi \ell^2 = L_xL_y (eB/\pi\hbar) $.
This expansion leads to the following expression for the
Ginzburg-Landau model
free energy:
\begin{eqnarray}
   \int f[\Psi] d\vec r & = & \{ \alpha_H \sum_p
\vert C_p \vert^2 + \frac{\beta}{2} \sum_{p_1p_2p_3p_4}
(L_y)^{-1}(2\pi \ell^2)^{-1/2}
\exp \lbrace -\frac{\ell^2}{2}[\sum_{i = 1}^{4}p_i^2-\frac{1}{4}
(\sum_{i = 1}^4 p_i)^2] \rbrace \nonumber \\
  &    &    \hspace{2.2in} \times \bar C_{p_1}
 \bar C_{p_2} C_{p_3} C_{p_4} \delta_{p_1+p_2, p_3+p_4} \}
\label{equ:3}
\end{eqnarray}
where $\alpha_H = \alpha(T) ( 1 - H / H_{c2}(T)) $.  Fluctuation
effects in the model are regulated by the dimensionless parameter,
$g \equiv \alpha_H (\pi \ell^2 / \beta k_B T)^{1/2}$.

The central quantity in our work is the superfluid-density
spatial correlation function which we define by
\begin{equation}
\chi_{SFD} (\vec R) \equiv \langle |\Psi(\vec r)|^2 |\Psi (\vec r
+ \vec
R) |^2 \rangle - \langle |\Psi ( \vec r ) |^2 \rangle \langle
|\Psi(\vec r + \vec R)|^2 \rangle
\label{final:1}
\end{equation}
Translational invariance of the system guarantees that the right hand
side of Eq.~(\ref{final:1}) is independent of $\vec r$.
The modulation of the average superfluid density when the
homogeneous system is disturbed by weak pinning can be expressed
in terms of
$\chi_{SFD} (\vec R)$:
\begin{equation}
\delta \langle |\Psi(\vec r)|^2 \rangle
= { -1 \over k_B T} \int d \vec r^\prime \chi_{SFD} (\vec r^\prime
- \vec r)
\delta \alpha (\vec r^\prime)
\label{last:1}
\end{equation}
where $\delta \alpha ( \vec r^\prime)$ reflects the modulation of the
mean-field transition temperature by pinning.
The Fourier transform of $\chi_{SFD}(\vec r)$ is
\begin{equation}
\chi_{SFD}(\vec q) \equiv
{ 1 \over L_x L_y } \int d^2 \vec r \int d^2 \vec r^\prime
[\langle |\Psi(\vec r)|^2 |\Psi(\vec r^\prime|^2 \rangle -
\langle |\Psi(\vec r|^2 \rangle \langle |\Psi(\vec r')|^2 \rangle ]
\exp [i \vec q \cdot ( \vec r - \vec r^\prime) ]
\label{eq:may1}
\end{equation}
We evaluate $\chi_{SFD}(\vec q)$ by expressing it in terms of
\begin{equation}
\Delta (\vec q) \equiv
\frac{1}{N_\phi}\sum_{p_1p_2}\bar C_{p_1} C_{p_2}
\delta_{p_2, p_1+q_y}
\exp [-i\ell^2q_x (p_1+p_2)/2]
\label{eq:may2}
\end{equation}
so that
\begin{equation}
\chi_{SFD}(\vec q) = {N_{\phi}^2  \over L_x L_y }
[\exp [- q^2 \ell^2 / 2 ]
\langle |\Delta (\vec q)|^2 \rangle - \langle \Delta (\vec q = 0)
\rangle^2 ]
\label{eq:may3}
\end{equation}

$\Delta(\vec q)$ satisfies the following informative
sum rule\cite{ref:4} for each configuration of the Ginzburg-Landau
system,
\begin{equation}
{ 1 \over N_{\phi}} \sum_{\vec q}
[\vert \tilde \Delta (\vec q) \vert^2 -1/N_{\phi}] = 0
\label{equ:5}
\end{equation}
where $\tilde \Delta (\vec q) \equiv \Delta (\vec q) / \Delta_0$ and
$\Delta_0 \equiv \Delta (\vec q = 0)$ is proportional to the
integrated superfluid density\cite{caveat}.
(Note that $\tilde \Delta (\vec q)$ is invariant
when the order parameter is multiplied by an overall constant.)
$\langle \vert \tilde \Delta (\vec q) \vert^2 \rangle$
is a particularly revealing quantity to examine in studying correlations
in the vortex liquid.  Eq.~(\ref{equ:5}) guarantees that (for large
$N_{\phi}$)
$\lim_{q \to \infty} N_{\phi} \vert \tilde \Delta (\vec q) \vert^2 =1$
for any vortex liquid configuration.  For example, in the high temperature
(vortex gas) limit $\langle \vert \tilde \Delta (\vec q) \vert^2 \rangle
=1/N_{\phi}$ for all $\vec q \ne 0$.
On the other hand it is easy to show that in the low temperature limit
the mean-field Abrikosov lattice configuration of the order parameter
gives
\begin{equation}
      \vert \tilde \Delta (\vec q) \vert^2 =\left\{
	\begin{array}{cc}
    1 & \mbox{   at $\vec q = \vec G$,} \\
    0 & \mbox{   otherwise.}
    \end{array} \right\}
\end{equation}
where $G$ is any reciprocal lattice vector.  Thus $N_{\phi} \langle
\vert \tilde \Delta (\vec q) \vert^2 \rangle \equiv s_V(q)$
shows exactly the behavior which would be expected for the static
structure
function of a classical fluid with $N_{\phi}$ particles in both
low temperature (solid) and high temperature (gas) limits.
It seems clear that $s_V(q)$ must be closely related to the
static structure factor of the zeroes of the order parameter,
the vortices, although we do not believe that they are identical at
all temperatures.  For the purposes of the present study it is
sufficient to observe that $h_V(q) \equiv s_V(q) - 1$, which we
call the vortex correlation function, is a convenient measure of of the
degree of correlation in this system.

\section{High Temperature Perturbation Expansion}

$\vert \Delta (\vec q) \vert^2$ can be expressed in the form
\begin{equation}
\vert \Delta (\vec q) \vert^2 = N_\phi^{-2}\sum_{p_1p_2p_3p_4}
\bar C_{p_1}
\bar C_{p_2}C_{p_3}C_{p_4}\delta_{p_1+p_2,p_3+p_4}(\delta_{p_1,p_4+q_y}
e^{i\ell^2q_x(p_4-p_2)}).
\end{equation}
At high temperatures we can evaluate its thermal average
\begin{equation}
\langle \vert \Delta (\vec q) \vert^2 \rangle \equiv  \frac{1}{Z}\int
\prod_{p} d\bar C_p dC_p
\vert \Delta (\vec q) \vert^2 e^{-\frac{\alpha_H}{K_B T} \int d\vec r
\vert
\Psi \vert^2} \times e^{-\frac{\beta}{2K_B T}\int d\vec r
\vert \Psi \vert^4}
\label{equ:4}
\end{equation}
where $Z$ is the partition function
\begin{equation}
 Z = \int \prod_{p} d\bar C_p dC_p
e^{-\frac{\alpha_H}{K_B T} \int d\vec r \vert
\Psi \vert^2} \times e^{-\frac{\beta}{2K_B T}\int d\vec r
\vert \Psi \vert^4}
\end{equation}
by expanding the contribution to thermal weighting factors from the
quartic contribution to the Landau-Ginzburg free energy.
The perturbation series can most easily be handled in terms of Feynman
diagrams~\cite{ref:1,parisi}. At $n$-th order , there are
contributions from diagrams with $(n+1)$ vertices and $2(n+1)$ edges
in which the edges represent the Gaussian approximation correlation
functions, $n$ vertices represent $|\Psi|^4$ terms
proportional to $\beta$ and the additional `external' vertex corresponds
to $\vert \Delta (\vec q) \vert^2$.  An important consequence of the
Landau quantization of order parameter fluctuations is the fact that
the Gaussian approximation correlation functions,
\begin{equation}
\langle \bar C_{p'} C_p \rangle = \delta_{p',p} k_B T / \alpha_H,
\label{final:2}
\end{equation}
are independent of the momentum $p$.
Each vertex has two directed outgoing lines
to represent the factors of $C_p$ associated with it and two incoming
lines to represent the factors of $\bar C_p$.  All diagrams containing
single-loop dressings of edges or vertices (except the external vertex)
can
be eliminated by following Ruggeri and Thouless~\cite{ref:1} and
expanding in terms of `Hartree-Fock' approximation correlation
functions.  This renormalization replaces\cite{ref:1} $\alpha_H$
in Eq.~(\ref{final:2}) by
$\tilde \alpha  \equiv  - 2 \alpha_H /(g^2 \pm [g^4+4 g^2]^{1/2})$.
(The $+$ ($-$) sign applies for $g<0$ ($g>0$).)  $\tilde \alpha$ remains
positive for all temperatures so that the renormalized
expansion parameter of the high-temperature perturbation
expansion, $x \equiv (\beta k_B T)/
(4 \pi \ell^2 \tilde \alpha^2)$, remains finite at all temperatures.
($\tilde \alpha \approx \alpha_H$ for $g \gg 0$,  $\tilde \alpha \approx
 |\alpha_H|/g^2$ for $g \ll 0$;
$\tilde \alpha = \alpha_H (1-4x)^{-1}$,  $g^2=(4x-1)^2/4x $ and $x=1/4$ at
$T_c^{MF}$.)
When the expansion is performed in terms of the
Hartree-Fock correlation functions
$\langle \vert \Delta (\vec q) \vert^2 \rangle $
is the sum of all the connected diagrams
without single-loop dressings except at the external vertex.

It is convenient to label the external vertex as vertex $1$ and to
label the incoming momenta at vertex $i$ as
$p_{2i}$ and $p_{2i-1}$.  Then the contribution to
$\langle | \Delta (\vec q) |^2 \rangle$
from an $n$-th order diagram is given by
$(k_B T / \tilde \alpha)^2 (-x)^n
N_{\phi}^{-1} I(\vec q)/n!$ where
\begin{eqnarray}
I (\vec q) \equiv & e^{-il^2q_y(q_x+iq_y)} (\frac{2 \pi}{\ell^2})^{-n/2}
\sum_{p_2} \int dp_3 ...\int dp_{2n+2} \prod_{\mu=2}^{n+1}
\delta(\sum_i(M_{\mu i}-N_{\mu i})p_i) \nonumber \\
&  \times (\delta_{ p_1^\prime+q_y}+ \delta_{p_2' +q_y})
e^{-i\ell^2(q_x+iq_y)p_2} \exp\{-\frac{\ell^2}{2}
\sum_{\mu =2}^{n+1}(p_{2\mu}-p_{2\mu -1})^2\},
\label{equ:12}
\end{eqnarray}
$p_1'$ and $p_2'$ are the two outgoing momentum labels at vertex
$1$, $\mu$ labels the vertices, and $M_{\mu i}$ is unity if $i$ goes into
the vertex $\mu$ and is zero otherwise, while $N_{\mu i}$ is unity if $i$
comes out of $\mu$ and is zero otherwise.\cite{ref:1}  To obtain this
result we have noted that the integral is invariant under a shift of all
momenta, set $p_1=0$ and multiplied by $N_{\phi}$.
It turns out that the integral $I(\vec q)$ can be evaluated exactly
and expressed in terms of the number of Euler paths
in the two subgraphs obtained by
deleting the external vertex and making the possible contractions.  The
two contractions correspond to the two delta functions which fix
either $p_1^{\prime}$ or $p_2^{\prime}$ in Eq.~(\ref{equ:12}).  We
denote the Euler path numbers for the two subgraphs by
$T^A$ and $T^B$.  The following result
is derived in the Appendix:
\begin{equation}
I ( \vec q) = \{\frac{1}{T^A}\exp(-\frac{T^B}{2T^A}q^2\ell^2)+
\frac{1}{T^B}\exp(-\frac{T^A}{2T^B}q^2\ell^2)\}
\label{int2}
\end{equation}
if $T^A \times T^B \neq 0$, and
\begin{equation}
I (\vec q) = (N_{\phi} \delta_{\vec q , 0} + 1)/ (T^A+T^B)
\label{int3}
\end{equation}
if $T^A \times T^B = 0$.

Using the above result for $I(\vec q)$ and writing the number
of appearances of a given graph as $4^{n+1}n!/G_{n+1,g}$ we
obtain the following formally exact expression for
$\langle \vert \Delta (\vec q ) \vert^2 \rangle  $:
\begin{eqnarray}
\langle \vert \Delta (\vec q) \vert^2 \rangle & =
& \frac{1}{N_\phi}(\frac{k_BT}{\tilde
{\alpha}})^2 \sum_{n= 0}^{\infty} \sum_g a_{n,g}(q)(-4 x)^n \nonumber \\
& = & \frac{1}{N_\phi}(\frac{k_BT}{\tilde {\alpha}})^2 (1+N_\phi
\delta_{\vec q, 0} - 4x\exp [-\frac{q^2\ell^2}{2}] +
                    4x^2\exp [-\frac{q^2\ell^2}{2}]\nonumber \\
& & +8x^2(1+N_\phi\delta_{\vec q, 0}) +
16x^2(0.5\exp [-\frac{q^2\ell^2}{4}]+\exp [-q^2\ell^2])+...),
\label{ptfinal}
\end{eqnarray}
where $g$ labels graphs and the sum at $n$-th order is over
($n+1$)-vertex graphs.  $a_{n,g}(q)$ is given by
\begin{equation}
a_{n,g}(q) = \left \{ \begin{array}{ll}
                  \frac{2}{G_{n+1,g}}(\frac{1}{T^A_{n,g}}
\exp [-\frac{T^B_{n,g}}{2T^A_{n,g}}q^2\ell^2] +
\frac{1}{T^B_{n,g}}\exp [-\frac{T^A_{n,g}}{2T^B_{n,g}}q^2\ell^2] ) &
\mbox{    If $T^A_{n,g} \times T^B_{n,g} \neq 0 $,} \\
                  \frac{2}{G_{n+1,g}(T^A_{n,g}+T^B_{n,g})}
(1 + N_\phi \delta_{\vec q , 0}) &
                      \mbox{    If $ T^A_{n,g} \times T^B_{n,g} = 0$,}
              \end{array} \right.
\label{graph}
\end{equation}
where $T^A_{n,g}$ and $T^B_{n,g}$ are the number of the Euler paths of
the two contracted $n$-vertex graphs and $G_{n+1,g}$
is the number of automorphisms of the $(n+1)$-vertex
graph with one external vertex.
The diagrams which appear up to second order in the series
and their associated properties are listed in Table~\ref{table:1}.
The explicit expression in Eq.~(\ref{ptfinal}) can be confirmed from
the entries in this Table.

We observe in Eq.~(\ref{graph}) that contributions which survive to
the large $|\vec q|$ limit come only from graphs where $T^A_{n,g}$
equals zero or $T^B_{n,g}$ equals zero and that the only terms in
$\langle |\Delta (\vec q)|^2 \rangle $ which are independent of
$N_{\phi}$ come from the same set of diagrams.  The contribution
independent of $N_{\phi}$ is
$\langle \Delta_0^2 \rangle \delta_{\vec q,0}$.
The remaining contributions to $\langle |\Delta (\vec q) |^2
\rangle $, which are proportional to $N_{\phi}^{-1}$,
contribute to $\chi_{SFD}(\vec q)$ and are due to
correlations in the thermally fluctuating superfluid density.
We thus obtain explicitly from the perturbation expansion that
\begin{equation}
\lim_{|\vec q| \to \infty} \langle |\Delta (\vec q) |^2 \rangle =
\langle \Delta_0^2 \rangle / N_{\phi}.
\label{largeq}
\end{equation}
This result was claimed earlier on the basis of the sum rule
(Eq.~(\ref{equ:5})).

The real space correlation function is given by
\begin{eqnarray}
\langle |\Psi (\vec r)|^2 |\Psi (\vec r
+ \vec R)|^2 \rangle
& = & \frac{1}{(2\pi )^2} \int d\vec q \langle |\Delta (\vec q)|^2 \rangle
 \exp [-q^2\ell^2/2] \exp [-i\vec q \cdot \vec R] \\
& = & (\frac{N_\phi}{L_xL_y})^2(\frac{k_BT}{\tilde{\alpha}})^2 \sum_{n= 0}
\sum_g a_{n,g}(R)(-4x)^n;
\end{eqnarray}
where
\begin{equation}
a_{n,g}(R) = \frac{2}{G_{n+1,g}T_{n+1,g}}
(\exp [-\frac{T^A_{n,g}}{2T_{n+1,g}}(\frac{R}{\ell})^2]+
\exp [-\frac{T^B_{n,g}}{2T_{n+1,g}}(\frac{R}{\ell})^2]).
\label{final:3}
\end{equation}
In Eq.(\ref{final:3}) $ T_{n+1,g} = T^A_{n,g}+T^B_{n,g}$ is the number
of Euler paths in the uncontracted ($n+1$)-vertex graph.  We see
again here that only graphs with $T_{n,g}^A=0$ or $T_{n,g}^B = 0$
remain finite for $ R \to \infty$ where correlations
vanish.  The contribution of this subset of graphs is
$\langle |\Psi (\vec r) |^2 \rangle^2$.

\section{Finite Order Results and Extrapolation to Low Temperatures}

We have written a computer program which generates all relevant graphs
represented by their adjacency matrices.~\cite{ref:6} From the adjacency
matrices we calculate the number of Euler paths for the
subgraphs, ($T^A_{n,g}$ and  $T^B_{n,g}$) and from the graph
generating algorithm we calculate the symmetry factor $G_{n+1,g}$.
In Fig.~(\ref{fig:1}), we show one of the
graphs which appears at second order in the expansion,
the two graphs which result from the deletion of its external vertex,
and the adjacency matrices of all three graphs.
(The number of Euler paths equals the determinant of the minor of matrix
appropriately formed from the corresponding adjacency matrix.~\cite{ref:6})
In this way, we have evaluated the series exactly
up to tenth order\cite{sixthorder}.
We have checked our results for
$\langle \vert \Delta (\vec q) \vert^2 \rangle $
by confirming that the sum rule
Eq.~(\ref{equ:5}), is satisfied and that the results for both
$\langle \vert \Delta_0 \vert^2 \rangle $ and
$\sum_{\vec q}\vert \Delta (\vec q)
\vert^2 \exp\{-\frac{1}{2}q^2\ell^2\}$ are correct order by order
in perturbation theory.  The latter two quantities can be
related to derivatives of the free energy.  The free energy is given by
\begin{equation}
F=-k_B T \ln Z = k_B T N_{\phi} ( \ln (\tilde \alpha / \pi k_B T)
+ f_{2D}(x))
\label{freen}
\end{equation}
where the perturbation expansion for $f_{2D}(x)$ was
first calculated by Ruggeri and Thouless~\cite{ref:1} and the expansion
was extended to order $x^{11}$ by Brezin, Fujita and Hikami.~\cite{ref:2}
Differentiating Eq.~(\ref{freen}) once with respect to $\beta$ and
find that
\begin{equation}
\sum_{\vec q} \langle \vert \Delta (\vec q)\vert^2 \rangle
\exp\{-\frac{1}{2}q^2l^2\}
= (\frac{k_BT}{\tilde{\alpha}})^2\times \frac{4 + (1-4x)
f^\prime_{2D}(x)}{1+4x}
\end{equation}
Similarly differentiating twice with respect to $\alpha_H$ gives
\begin{equation}
\langle \vert \Delta_0 \vert^2 \rangle - \langle \Delta_0
\rangle^2 = \frac{1}{N_\phi}(\frac{k_BT}{\tilde{\alpha}})^2 ( \frac{(1-4x)
(1-2xf^\prime_{2D}(x))}{(1+4x)^3} - \frac{4x(f^\prime_{2D}(x)
+xf_{2D}^{\prime \prime})}{(1+4x)^2})
\end{equation}
where $\langle \Delta_0 \rangle  = (k_BT/\tilde{\alpha})
(1-2xf^\prime_{2D}(x))/(1+4x)$.

The asymptotic high-temperature expansion of
$\langle \vert \Delta (\vec q) \vert^2 \rangle $
can be extrapolated to low temperatures using Pad\'{e}
approximants to describe the $x$ dependence at
each value of $q$.  Comparisons between Pad\'{e} approximants of
the series and our Monte Carlo simulation results~\cite{ref:4} are
given in Fig.~\ref{fig:2} for different
temperatures.  $s_V(q)$ shows a well-defined peak for $q$ near $|\vec G|$,
where $\vec G$ is a reciprocal lattice vectors of the vortex solid,
Such a peak is characteristic of a strongly correlated liquid.
At low temperatures the [2n, 2n] Pad\'{e} approximants appear
to overestimate the correlations while the [2n+1,2n+1] approximants
appears to underestimate the correlations.  Quantitative agreement
with Monte Carlo simulation results is obtained for
temperatures above $T_c^{MF}$ and
for temperatures  below $T_c^{MF}$ with $g^2 < 5$.
A continuous phase-transition to a vortex solid state
at low temperatures would be indicated by the divergence of
$s_v(q)$ at $q$ near $|\vec G|$ and hence by a pole in the
Pad\'{e} approximant to $\langle \vert \Delta (\vec q) \vert^2 \rangle$
along the positive real $x$ axis.  It might be significant that
for the [4,4] Pad\'{e} approximant poles on the positive real axis appear
over a finite range of $q$ near $|\vec G|$.
In Fig.~\ref{fig:3} we plot the $q$ dependence of the
value of $x_c$ at which these poles occur.  The highest temperature
at which a pole occurs is $x= 9.69 $ corresponding to $g^2 = 36.8 $
for $q = 0.91 |\vec G|$.
This value of $g^2$ compares with the value $g^2= 43.5 \pm 1.0 $ at
which a weak first order phase transition occurs according to the
Monte Carlo calculations.~\cite{ref:4}  No poles on the positive real
axis for $x$ occur
in the $[5,5]$ Pad\'{e} but we expect that they will recur at slightly
lower temperatures in the $[6,6]$ Pad\'{e}.
We believe that the structure in $s_V(q)$ is a precursor
of the solidification of the vortex lattice at low temperatures.
The proximity of the vortex lattice state is apparent in the
perturbation theory for $s_V(q)$ but is hidden in the
perturbation theory for the free energy because of the weakness of the
thermodynamic singularity associated with the phase transition.
However, accurate estimates of the transition temperature based on
the perturbation expansion for $s_V(q)$ would require
calculations to be carried out to higher order than we have found
possible to date.

\section{Summary and Conclusions}

We have studied the superfluid density spatial correlation
function $\langle \vert \Delta (\vec q) \vert^2\rangle $ using high
temperature
perturbation expansions and extrapolated our results
to temperatures below the mean-field transition temperature by means of
Pad\'{e} approximants.  Good agreements with Monte Carlo
simulation data is obtained for $g > -\sqrt{5}$.
Our results demonstrate that the vortex liquid is strongly
correlated below $T_c^{MF}$.   A result argued for previously
on the basis of a sum rule which related the large wavevector
limit of the correlation function to the average superfluid
density was obtained explicitly from the perturbation
expansion.  We argue that these perturbation
expansion studies bespeak the phase transition to a two-dimensional
vortex solid that is believed to occur at lower temperatures.
However, higher order calculations than we have been able to
complete to date would be necessary in order to obtain estimates
of the transition temperature which are competitive in
accuracy and reliability with those obtained previously from
Monte Carlo simulations of the same model.
This work was supported by the Midwest
Superconductivity Consortium through D.O.E.
grant no. DE-FG-02-90ER45427.   The authors are grateful to
Steve Girvin, Brendan McKay  and Lian Zheng for helpful interactions.

\appendix
\section*{}
The integral to be evaluated has the form:

\begin{eqnarray}
I (\vec q) \equiv & e^{-il^2q_y(q_x+iq_y)} (\frac{2 \pi}{\ell^2})^{-n/2}
\sum_{p_2} \int dp_3 ...\int dp_{2n+2} \prod_{\mu=2}^{n+1}
\delta(\sum_i(M_{\mu i}-N_{\mu i})p_i) \nonumber \\
&  \times (\delta_{ p_1^\prime+q_y}+ \delta_{p_2' +q_y})
e^{-i\ell^2(q_x+iq_y)p_2} \exp\{-\frac{\ell^2}{2}
\sum_{\mu =2}^{n+1}(p_{2\mu}-p_{2\mu -1})^2\},
\label{equ:12a}
\end{eqnarray}

Of the $(2n+1)$ variables $p_i$ only $n$
are independent because of the $(n+1)$ delta functions.
It is convenient\cite{ref:5} to choose $s_{\mu}=p_{2\mu}-p_{2\mu -1}$
($\mu = 2, \cdots, n+1$)
as the $n$ independent variables.  We label the contributions to
$I(\vec q)$
from the two choices for the delta function at the external vertex as
$I^A(\vec q)$ and $I^B(\vec q)$ respectively.  We use this delta
function to
eliminate the sum over $p_2=s_1$ which can be expressed as a linear
function of the independent variables.  For $X$ =$A$ or $B$
\begin{equation}
	s^X_1 = - \alpha^X q_y + \sum_{\mu=2}^{n+1} t^X_\mu s_\mu.
\end{equation}
The integral over the independent variables can then be expressed
in terms of these coefficients.
The Jacobian for changing variables from $p_\mu$
to $s_{\mu > 1}$ is independent of $q_y$; at $ q_y=0 $ the change of
variables is identical to that required for the diagram obtained
by deleting the external vertex and contracting the two
outgoing edges at that vertex with the two incoming edges. (The
two ways of doing the contraction correspond to $I^A(\vec q)$ and
$I^B(\vec q)$ respectively.)  Following the
work of McCauley and Thouless\cite{ref:5} we note that the
inverses of the required Jacobians
equal the number of Euler paths\cite{ref:6} for the
$n$-vertex graphs which results from the deletion of the external vertex
and the two possible contractions.  We denote the Euler path
numbers by $T^X$.
The integral over the $s_{\mu}$ is then
elementary and we obtain
\begin{equation}
I^X (\vec q) =
(e^{i \ell^2 q_x q_y})^{n^X_1} (e^{ - \ell^2|q_y|^2})^{n^X_2}
(e^{ -\ell^2|q_x|^2})^{n^X_3}
\frac{1}{T^X}
\label{int1}
\end{equation}
where $n^X_1= -1+\alpha^X - \sum_{\mu=2}^{n+1} (t^X_{\mu})^2$,
$n^X_2=-1+\alpha^X-\frac{1}{2}\sum_{\mu=2}^{n+1} (t^X_{\mu})^2$ and
$n^X_3 = \frac{1}{2}\sum_{\mu=2}^{n+1} (t^X_{\mu})^2$.  However,
we know that
$I^X (\vec q)$ is real which leads to the requirement that $n^X_1=0$ and
implies that $I^X (\vec q) = \exp (- q^2 \ell^2 (\alpha^X -1)/2)/T^X$.
The value of $\alpha^X$ can be inferred by noting
that $\sum_q \exp ( -q^2 \ell^2 /2 ) I^X
(\vec q) $ is proportional to an integral which appears in the expansion
of the
free energy and equals $N_{\phi}/T$ where $T$ is the number of Euler
paths in the original graph before deletion.  It follows that
$\alpha^X= T/T^X$ and hence that
\begin{equation}
I ( \vec q) = \{\frac{1}{T^A}\exp(-\frac{T^B}{2T^A}q^2\ell^2)+
\frac{1}{T^B}\exp(-\frac{T^A}{2T^B}q^2\ell^2)\}.
\label{int2a}
\end{equation}
(Note that $T^A + T^B = T$.)
This analysis fails in the special case where
$T^A \times T^B = 0$.
For that case we have  $I^X(\vec q) = n^X\delta_{\vec q , 0}$ or
$I^X(\vec q)
= n^X$ and a similar analysis gives the corresponding $n^X = N_\phi / T$
or $n^X  = 1 / T$ and so
\begin{equation}
I (\vec q) = (N_{\phi} \delta_{\vec q , 0} + 1)/ (T^A+T^B).
\label{int3a}
\end{equation}

\begin{table}
\caption{All diagrams up to second order in the high-temperature
perturbation expansion. The open circle in
each diagram represents the external vertex.}
\label{table:1}
\end{table}

\begin{figure}
\caption{Example of the adjacency matrix representation of a graph
which appears at
second order in the expansion and of
the calculation of Euler path numbers of the graph.
The graph is represented by an adjacency matrix $A$ with elements
$a_{ij}$ where $a_{ij}$ is equal to the number of edges going from vertex
$i$ to vertex $j$.  The number of Euler paths is equal to the determinant
of the minor of the matrix $B$ with elements defined by
$b_{ij} = \delta_{ij}\sum_k a_{ik} - a_{ij}$.
The open circle in the uncontracted graphs represents the external vertex.
The two graphs on the left result from the two possible
contractions after deletion of the external vertex.
The contribution of this graph to the series is $a_2(q) =
0.5 \exp (-q^2\ell^2/4) + \exp (-q^2\ell^2)$.}
\label{fig:1}
\end{figure}

\begin{figure}
\caption{Comparisons of the perturbation calculation and
Monte Carlo simulations.
The $y$-axis label is $\langle \vert \tilde \Delta (\vec q) \vert^2
 \rangle  $, while the $x$-axis is the wave vector in the units
of the inverse magnetic length $\ell^{-1}$.
 The upper-left panel is for a temperature above $T_c^{MF}$,
 while the other panels are for temperatures below $T_c^{MF}$.
 The arrows in the plots indicate the location of reciprocal
 vector, $|\vec G|$, of the vortex lattice state.   The Pad\'{e}
 approximants sometimes behave poorly at small wavevectors.
 ($q\ell \sim 1.0$.)
 We have `clipped' anomalous small wavevector behavior in some
 of these figures so as not to obscure the behavior for $q$ near $G$.}
\label{fig:2}
\end{figure}

\begin{figure}
\caption{Wavevector dependence of one of the poles in the
	$[4,4]$ Pad\'{e} approximant.  The pole moves from
	negative to positive $x$ by passing through $\infty$.
      This approximant has poles for $q$ near $G$ at
	temperatures which are somewhat higher than the
	temperature where a (weakly) first order phase
	transition is indicated by Monte Carlo simulations.  (See text.)
      The arrows in the plot indicate the location of reciprocal
      vector $|\vec G|$ of the vortex lattice state. }
\label{fig:3}
\end{figure}


\begin{thebibliography}{10}

\bibitem{abrikosov} A.~A. Abrikosov, Zh. Eksp. Teor. Fiz. {\bf 32},
1442 (1957).

\bibitem{liquid} D.~R. Nelson, Phys. Rev. Lett. {\bf 60}, 1415 (1988);
P.~L. Gammel, L.~F. Schneemeyer, J.~V. Wasczak, and D.~J. Bishop,
Phys. Rev. Lett. {\bf 61}, 1666 (1988).

\bibitem{leeshenoy} P.~A. Lee and S.~R. Shenoy, Phys. Rev. Lett.
{\bf 28}, 1025 (1972).

\bibitem{ref:1}
G.~J.~Ruggeri and D.~J.~Thouless, J. Phys. F {\bf 6}, 2063 (1976).

\bibitem{ref:2}
E.~Br\'{e}zin, A.~Fujita, and S.~Hikami,
\newblock Phys. Rev. Lett. {\bf 65}, 1949 (1990);
S.~Hikami, A.~Fujita, and A.~I.~Larkin,
\newblock Phys. Rev. B {\bf 44}, 10400 (1991).

\bibitem{ref:3} Zlatko Te\v{s}anovi\'{c} and L.~Xing,
Phys. Rev. Lett. {\bf 67}, 2729 (1991).

\bibitem{kato}
Yusuke Kato and Naoto Nagaosa, Phys. Rev. B. {\bf 47}, 2932 (1993).

\bibitem{ref:4}
Jun Hu and A.~H. MacDonald,
\newblock Phys. Rev. Lett. {\bf 71}, 432 (1993).

\bibitem{moore} J.~A. O' Neill and M.~A. Moore,
\newblock Phys. Rev. B {\bf 48}, 374 (1993).

\bibitem{ajp} A.~H. MacDonald, Hiroshi Akera, and M.~R. Norman,
Aust. J. Phys. {\bf 46}, 333 (1993).

\bibitem{ikeda} The limits of validity of this approximation are
thoroughly discussed by Ryuske Ikeda, preprint (1993).

\bibitem{vecpot} We also neglect fluctuations in the vector potential.
This approximation is permitted as long as the temperature is not
too far below the mean-field transition temperature or the
films are thin and widely separated.

\bibitem{caveat} We remark that Eq.~(\ref{equ:5}) is exact
even for finite-size systems
with quasiperiodic boundary conditions applied to the order
parameter.  In that case the number of $\vec q$ values
in the sum over wavevectors is $N_{\phi}^2$.  Finite-size
effects have to be considered carefully when terms of
order $N_{\phi}^{-1}$ are retained on the left hand side of
Eq.~(\ref{equ:5}).

\bibitem{parisi} Giorgio Parisi, {\it Statistical Field Theory}
(Addison-Wesley, New York, 1988).

\bibitem{ref:5}
G.~P.~McCauley and D.~J.~Thouless,
\newblock J. Phys. F {\bf 6}, 109 (1976).

\bibitem{ref:6}
C.~W.~Marshall,
\newblock {\em Applied Graph Theory},
\newblock Wiley-Interscience, 1971.

\bibitem{sixthorder} The perturbation expansion of the Fourier transform
of this function has
previously been evaluated to sixth order using a different approach by
Ryuske Ikeda, Tetsuo Ohmi, and Toshihiko Tsuneto,
J. Phys. Soc. Jpn. {\bf 59}, 1397 (1990).

\end{thebibliography}
\end{document}